\documentclass[a4paper,10pt]{article}
\usepackage{jheppub}
\usepackage{hyperref}

\usepackage{graphicx}
\usepackage{dcolumn}
\usepackage{bm}

\usepackage{enumitem}
\usepackage{xcolor}
\definecolor{codepurple}{rgb}{0.58,0,0.82}

\usepackage[bottom]{footmisc}

\usepackage{float}
\usepackage{csquotes}
\usepackage{tensor}
\usepackage{mathtools}
\usepackage{physics}
\usepackage{cancel}

\colorlet{linkequation}{codepurple}
\newcommand*{\SavedEqref}{}
\let\SavedEqref\eqref
\renewcommand*{\eqref}[1]{%
  \begingroup
    \hypersetup{
      linkcolor=linkequation,
      linkbordercolor=linkequation,
    }%
    \SavedEqref{#1}%
  \endgroup
}

\newcommand{\Lagr}{\mathcal{L}}

\begin{document}

\title{TDiff in the Dark: \\Gravity with a scalar field invariant under transverse diffeomorphisms}

\author[a]{Darío Jaramillo-Garrido,}
\author[a,b]{ Antonio L. Maroto,}
\author[a,b]{ and Prado Martín-Moruno}
\affiliation[a]{Departamento de Física Teórica, Ciudad Universitaria, Universidad Complutense de Madrid, E-28040 Madrid, Spain}
\affiliation[b]{Instituto de Física de Partículas y del Cosmos (IPARCOS-UCM), Universidad Complutense de Madrid, E-28040 Madrid, Spain}

\emailAdd{djaramil@ucm.es}
\emailAdd{maroto@ucm.es}
\emailAdd{pradomm@ucm.es}

\date{\today}

\abstract{
    We reflect on the possibility of having a matter action that is invariant only under transverse diffeomorphisms. This possibility is particularly interesting for the dark sector, where no restrictions arise based on the weak equivalence principle. In order to implement this idea we consider a scalar field which couples to gravity minimally but via arbitrary functions of the metric determinant. We show that the energy-momentum tensor of the scalar field takes the perfect fluid form when its velocity vector is time-like. 
    We analyze the conservation of this tensor in detail, obtaining a seminal novel result for the energy density of this field in the kinetic dominated regime. Indeed, in this regime the fluid is always adiabatic and we obtain an explicit expression for the speed of sound. Furthermore, to get insight in the gravitational properties of these theories, we consider the fulfillment of the energy conditions, concluding that nontrivial physically reasonable matter violates the strong energy condition in the potential domination regime. On the other hand, we present some shift-symmetric models of particular interest. These are: constant equation of state models (which may provide us with a successful description of dark matter or dark radiation) and models presenting different gravitational domains (characterized by the focusing or possible defocusing of time-like geodesics), as it happens in unified dark matter-energy models.
}

\preprint{IPARCOS-UCM-23-064}

\maketitle

\section{Introduction}

 We have entered the era of precision cosmology and the observational data indicate that our Universe is currently expanding in an accelerated way \cite{riess_observational_1998,perlmutter_measurements_1999}. The standard cosmological model assumes that this acceleration is due to a cosmological constant. However, other possible explanation for this acceleration within the framework of General Relativity (GR) is based on a dynamical component referred to as dark energy \cite{Huterer:1998qv,planck_collaboration_planck_2016}. A simple way of modeling dynamical dark energy is using a scalar field with a canonical or non-canonical kinetic term, such as quintessence, $k$-essence or kinetic gravity braiding \cite{amendola_dark_nodate,Armendariz-Picon:2000ulo,Deffayet:2010qz}. 
 On the other hand, although up to the date GR remains our best description of gravity, it is also possible that the current accelerated expansion of the Universe indicates a breakdown of GR at cosmological scales. Therefore, in the last years there has been an increased interest in alternative theories of gravity (see, for example, \cite{CANTATA:2021ktz} for a review). For example, Horndeski theory \cite{Horndeski:1974wa,Deffayet:2011gz,Kobayashi:2011nu}, based on introducing a scalar field non-minimally coupled with gravity, has provided us with the perfect framework for investigating the implications of having an additional gravitational degree of freedom \cite{Martin-Moruno:2015bda,Kase:2018aps}.

Following this spirit one can also reconsider the symmetries playing a fundamental role in a gravitational context. In particular, in recent years, interest has grown in theories which present a broken Diffeomorphism (Diff) invariance \cite{Alvarez:2005iy,Alvarez:2006uu,maroto_tdiff_2023}. The most popular alternative is the so-called Unimodular Gravity (UG) \cite{Henneaux:1989zc,Unruh:1988in} (see, for instance, \cite{Carballo-Rubio:2022ofy} for a review), first proposed by Einstein in 1919. In UG, the metric determinant is taken to be a constant, non-dynamical field, and in this manner the invariance of GR under the full group of diffeomorphisms is broken down to only invariance under the more restrictive Transverse Diffeomorphisms (TDiffs) and, in addition, Weyl rescalings. Nevertheless, the focus in this work is placed on TDiffs, which are coordinate transformations with the determinant of the Jacobian matrix equal to unity, without assuming Weyl invariance. Infinitesimally, if we consider a coordinate transformation $x^\mu \rightarrow \hat{x}^\mu = x^\mu + \xi^\mu(x)$, then what we do is require that $\partial_\mu \xi^\mu = 0$ (hence the name ``transverse''). An immediate and important consequence of TDiff invariance is that we can no longer distinguish between tensors and tensor densities, since the Jacobian must equal 1. In particular, the metric determinant is a true scalar and, therefore, the function of the metric determinant appearing in the Lagrangian is no longer fixed to be $\sqrt{g}$ \cite{Alvarez:2009ga}.
Notice that is always possible to reformulate unimodular gravity in a  generally covariant fashion by introducing additional fields \cite{Henneaux:1989zc,Kuchar:1991xd}. Gravitational theories invariant under TDiffs but involving a dynamical metric determinant have also been considered in \cite{Pirogov:2009hr,Pirogov:2011iq} and their potential connections with the cosmological dark sector have been explored in  \cite{Pirogov:2005im,Pirogov:2015bqc}.

In this work we do not consider the breaking of Diff invariance in the gravitational action, which shall remain the unchanged Einstein-Hilbert action. Rather, the symmetry breaking from Diff to TDiff shall be taken to occur explicitly in the matter action (consequently, however, affecting the full theory). In particular, we take a scalar field with a Lagrangian density containing different functions of the metric determinant in the kinetic and in the potential parts \cite{Alvarez:2009ga,maroto_tdiff_2023}, in such a way that the field can still be considered minimally coupled to gravity.
In this framework, it is of course possible to consider TDiff invariant visible matter. In that case, the phenomenological viability of the models implies a particular relation between the coupling functions (in particular, they must be the same) \cite{Alvarez:2009ga,maroto_tdiff_2023}. Nevertheless, it is worthy to note that if the breaking from Diff to TDiff takes place only in the dark sector, there is no reason to assume any relation between those functions.

The main aim of the present work is to perform a general study which allows us to gain some intuition on the new phenomenology that can be described within this framework. 
The only assumption we shall make is that the field derivative is a time-like vector. As we will explicitly show in this case the energy-momentum tensor (EMT) of the scalar field takes the form of a perfect fluid (see, for example, \cite{Pujolas:2011he}, for an interesting field theory in which this is not the case). This will allow us to extend some of the results obtained in \cite{maroto_tdiff_2023} in the cosmological context to general geometries.  
On the other hand, it should be noted that the full Diff invariance of GR implies that the EMT is automatically conserved on the solutions to the equations of motion of the theory (this is shown in any GR textbook, see \cite{Velten:2021xxw} for a review). However, when Diff symmetry is broken down to TDiff only on the matter sector, this conservation is no longer an automatic consequence of the field equation, but it is still implied by the Bianchi identities. This imposes an additional constraint on the metric that shall be considered in detail. As we will conclude, this new constraint allows us to fix the physical component of the metric that one could no longer choose arbitrarily due to the symmetry lost.

The work is organized as follows. In section \ref{section: Preliminary concepts} we review some definitions and techniques, and present the TDiff scalar field model under consideration (together with two limiting cases that shall be studied throughout). In section \ref{section: The perfect fluid approach} we describe the scalar field as a perfect fluid. Its energy conditions are considered in section \ref{section: Energy conditions}. Section \ref{section: Conservation of the EMT} presents a detailed analysis of the EMT conservation in the potential and kinetic regimes, together with its consequences on the metric and on the properties of the fluid. 
In section \ref{section: Some particular cases} we present some TDiff models of particular interest, analyzing their gravitational consequences through the application of the energy conditions. 
Section \ref{section: Discussion and conclusions} is devoted to the main conclusions of the work. Finally, in Appendices \ref{section: Covariantized action} and \ref{section: Calculation of the transverse constraint} we include some additional information and calculations, which may be skipped without losing the thread of the discussion.

\section{Preliminary concepts}\label{section: Preliminary concepts}
\noindent We present in this section a short review of the definitions and techniques employed throughout the work, and introduce the model we shall study. Our conventions include units in which $\hbar=c=1$, the usage of metric signature $(+, -, -, -)$, and the notation $g = \abs{\det(g_{\mu\nu})}$.

\subsection{Presenting the model}
\noindent The total action we shall consider in this work is
\begin{equation}\label{eq: total action}
    S = S_\text{EH} + S_m \,,
\end{equation}
the gravitational action being the Einstein-Hilbert action
\begin{equation}\label{eq: E-H action}
    S_\text{EH} = - \frac{1}{16\pi G} \int d^4x\, \sqrt{g} \, R
\end{equation}
($R = g^{\mu\nu} R_{\mu\nu}$ the Ricci scalar), and the matter action taken to be of the form
\begin{equation}
    S_m = \int d^4x\, \tilde{\Lagr}_m \,.
\end{equation}
In the above expression we find the so-called Lagrangian density $\tilde{\Lagr}_m = \tilde{\Lagr}_m(\Psi,\partial_\mu\Psi,g_{\mu\nu})$, which depends on the matter fields $\Psi$, their first derivatives $\partial_\mu\Psi$, and the metric $g_{\mu\nu}$. We remark that $\tilde{\Lagr}_m$ must be a TDiff scalar since the volume element $d^4x$ is invariant under TDiffs.

Applying to the total action \eqref{eq: total action} the stationary action principle with respect to variations in the spacetime metric yields the Einstein field equations
\begin{equation}
    G_{\mu\nu} = R_{\mu\nu} - \frac{1}{2}R g_{\mu\nu} = 8\pi G \, T_{\mu\nu} \,.
\end{equation}
We find in these equations the Energy-Momentum Tensor (EMT), whose definition is the usual one in GR
\begin{equation}\label{eq: def EMT}
    T_{\mu\nu} = \frac{2}{\sqrt{g}} \frac{\delta S_m}{\delta g^{\mu\nu}} \,.
\end{equation}
One may worry about whether this definition extends to a situation in which we are explicitly changing the matter action with respect to that in GR, as we shall see in this work, but in fact it makes sense since we will not be altering the gravitational sector (indeed, we use $S_\text{EH}$). In other words, we are not altering the left hand side of Einstein's equations, so we want whatever comes out on the other side to be associated to the EMT in the usual manner. As a final note regarding EMTs, since it will be useful throughout the work, we also recall here that for a perfect fluid
\begin{equation}\label{eq: EMT perfect fluid}
    T_{\mu\nu} = (\rho + p)u_\mu u_\nu - pg_{\mu\nu} \,,
\end{equation}
with $\rho$ the energy density of the fluid, $p$ its pressure, and $u^\mu$ represents a time-like unit vector field ($u^2 \equiv u_\mu u^\mu = 1$) which we interpret as the velocity field of the fluid.
Before presenting the model, let us recall the following relation for a given vector field $V^\mu$:
\begin{equation}\label{eq: divergence nabla and partial}
    \nabla_\mu V^\mu = \frac{1}{\sqrt{g}} \partial_\mu \left( \sqrt{g} \, V^\mu \right) \,,
\end{equation}
which shall be of use later on. We also remark here that throughout the work the action of the covariant derivative on (TDiff) tensors maintains the usual definitions.\\
\indent The model we shall study in this work is that of a scalar field $\psi(x)$ in which the kinetic and potential terms are coupled not only differently than in GR, but also differently from each other. The matter action reads \cite{Alvarez:2009ga}
\begin{equation}\label{eq: model matter action}
    S_m = \int d^4x\, \left\{ \frac{f_k(g)}{2} g^{\mu\nu} \partial_\mu \psi \partial_\nu \psi - f_v(g) V(\psi) \right\} \,.
\end{equation}
Here, $f_k(g)$ and $f_v(g)$ are arbitrary functions of (the absolute value of) the metric determinant $g=\abs{\det(g_{\mu\nu})}$, and the subscripts make reference to the kinetic and potential terms, respectively. The GR limit, in particular, would correspond to $f_k, f_v \propto \sqrt{g}$. It is interesting to note that even though the couplings are not the usual they are still minimal, meaning that there is no coupling of the field to the curvature (second derivatives of the metric). We also remark that the matter action \eqref{eq: model matter action} is in general not invariant under the full group of diffeomorphisms (Diff invariance is only restored in the GR limit we mentioned above, as it is easy to verify). Nevertheless, a moment's reflection reveals that our model will be invariant under the reduced group of TDiff symmetries, since in that case the Jacobian is unity and the metric determinant is a scalar, together with functions of it. The Euler-Lagrange equation of motion (EoM) for the scalar field reads
\begin{equation}\label{eq: complete EoM for psi}
    \partial_\mu\big( f_k(g) \partial^\mu \psi \big) + f_v(g) V'(\psi) = 0 \,,
\end{equation}
where $\partial^\mu \psi = g^{\mu\nu} \partial_\nu\psi$ and $V'(\psi) = dV / d\psi$ (in general, a prime on a function will denote differentiation with respect to its argument). Also, using definition \eqref{eq: def EMT}, the associated EMT turns out to be
\begin{equation}\label{eq: model EMT}
    T_{\mu\nu} = \frac{2}{\sqrt{g}} \left\{ \frac{1}{2} f_k(g) \partial_\mu \psi \partial_\nu \psi + g \left[ f_v'(g) V(\psi) - \frac{1}{2} f'_k(g) \partial_\alpha \psi \partial^\alpha \psi \right] g_{\mu\nu} \right\} \,,
\end{equation}
which is equivalent to the one presented in reference \cite{maroto_tdiff_2023}.

Before proceeding any further, it is interesting to note that it is possible to rewrite the model action with broken Diff invariance in a Diff invariant way via the introduction of additional fields \cite{maroto_tdiff_2023}. The interested reader is directed to Appendix \ref{section: Covariantized action} for a short discussion on this regard.

We shall now present a couple of simple limiting regimes of our model which will be studied throughout the work: potential domination and kinetic domination. These are interesting not only because they greatly simplify the treatment, but also because they help us in gaining intuition about the underlying physics, which we may then use. On the one hand, possible applications of a dominant kinetic term include shift-symmetric models for dark energy \cite{Germani:2016gzh} (which are actually purely kinetic), as well as cases in which the field is rapidly changing and the kinetic term dominates over the potential (e.g. in fast-roll scenarios at the end of inflation). On the other hand, the dominant potential behavior may be found in cases where the field is slowly varying (e.g. slow-roll inflation) or, with a dark sector application in mind, it also allows us to study dark energy models.

\subsubsection{Potential domination}
\noindent In the potential domination limit we effectively neglect the kinetic term, so that
\begin{equation}
    S_m \simeq \int d^4x\, \left[-f_v(g)\,V(\psi)\right]
\end{equation}
and the EoM simplify to
\begin{equation}
    f_v V' = 0 \, \implies \, V' = 0\,,
\end{equation}
wherever $f_v \neq 0$. The fact that $V'=0$ implies in turn that the field takes on the constant value $\psi=\psi_0$ which is the extremum of the potential, i.e. $V'(\psi_0) = 0$, and as a result 
\begin{equation}
    V(\psi) = V(\psi_0) \equiv V_0 = \text{const.}
\end{equation}
Finally, we see that within this dominant potential approximation the EMT is written as
\begin{equation}\label{eq: EMT in potential dom}
    T_{\mu\nu} = 2\sqrt{g}\,f'_v V g_{\mu\nu} \,,
\end{equation}
so it is proportional to the metric.

\subsubsection{Kinetic domination}
\noindent In the dominant kinetic limit, we effectively neglect the potential term. Thus,
\begin{equation}
    S_m \simeq \int d^4x\,  \frac{f_k(g)}{2} g^{\mu\nu} \partial_\mu \psi \partial_\nu \psi
\end{equation}
and the EoM become
\begin{equation}\label{eq: EoM kinetic}
    \partial_\mu \left(f_k \partial^\mu \psi\right) = 0\,.
\end{equation}
Let us note that, in this limit, the EoM essentially reveals the existence of a conserved current $J^\mu = \frac{f_k}{\sqrt{g}} \, \partial^\mu \psi$ such that $\nabla_\mu J^\mu = 0$ according to relation \eqref{eq: divergence nabla and partial}. This is not surprising since the kinetic limit, in the way in which we are treating it, presents shift symmetry, and this is the corresponding conserved current \cite{Kobayashi:2011nu}. Note also that for GR we recover $J^\mu=\partial^\mu\psi$, as we should since for a massless scalar field one has $\Box\psi=0$. Finally, within the kinetic approximation the EMT reads
\begin{equation}
    T_{\mu\nu} = \frac{1}{\sqrt{g}} \left[ f_k \partial_\mu \psi \partial_\nu \psi - g f'_k (\partial\psi)^2 g_{\mu\nu} \right] \,,
\end{equation}
where $(\partial\psi)^2 = \partial_\alpha \psi \partial^\alpha \psi$.

\subsection{Review of Energy Conditions}
\noindent The so-called Energy Conditions (ECs) are a set of requirements that one imposes on the EMT upon arguing that they are ``physically reasonable'' \cite{hawking_large_2008,Martin-Moruno:2017exc,Bouhmadi-Lopez:2019zvz}. The most widely used are the Null, the Weak, the Strong, and the Dominant Energy Conditions (NEC, WEC, SEC, and DEC, respectively). Given any time-like and null vectors $v^\mu$ and $k^\mu$, respectively, these energy conditions can be formulated in the following way \cite{hawking_large_2008,poisson_relativist39s_2004,Martin-Moruno:2017exc,Bouhmadi-Lopez:2019zvz}:

\begin{itemize}
    \item DEC: The energy density measured by any observer is non-negative and propagates in a causal way. That is 
    \begin{equation}\label{NEC}
        T_{\mu\nu} v^\mu v^\nu \geq 0\,,\quad\text{and}\quad F^\mu \equiv -T^{\mu\nu} v_\nu\quad \text{is causal}.
    \end{equation}
    \item WEC: The energy density measured by any observer has to be non-negative. So 
    \begin{equation}\label{WEC}
        T_{\mu\nu} v^\mu v^\nu \geq 0\,.
    \end{equation}
    \item SEC: Gravity is always attractive in GR. Taking into account the Raychaudhuri equation \cite{Kar:2006ms}, this implies
    \begin{equation}\label{SEC}
        \left(T_{\mu\nu} - \frac{1}{2}T\indices{^\alpha_\alpha} g_{\mu\nu}\right) v^\mu v^\nu \geq 0 \,.
    \end{equation}
    \item NEC: The SEC and the WEC are satisfied in the limit of null observers. Both conditions in that limit imply
    \begin{equation}
        T_{\mu\nu} k^\mu k^\nu \geq 0 \,.
    \end{equation}
\end{itemize}
Therefore, the fulfillment of the SEC implies that the NEC is satisfied. On the other hand, the DEC implies the WEC that leads to the fulfillment of the NEC. Thus, violations of the NEC imply that all the other ECs are violated.

It should be noted that the only EC in which a theory of gravity is specified is in the SEC. Indeed, the SEC is just the Time-like Convergence Condition (TCC) taking into account GR. The TCC states that gravity is always attractive, which is based on requiring that $R_{\mu\nu} v^\mu v^\nu \geq 0$. This conditions comes from demanding the convergence of (vorticity-free) time-like geodesics in the Raychaudhuri equation \cite{Martin-Moruno:2017exc}. Nowadays, it is well-known that the TCC (and, therefore, the SEC in GR) should be violated during the cosmic phases of accelerated expansion. Therefore, in the present work we should consider that the WEC and DEC should be satisfied by physically reasonable classical matter, whereas we would consider the possible violation of the TCC as a potential signal of the dark sector. On the other hand, since in the gravitational framework under investigation in the present paper the Einstein equations are satisfied, we will directly consider the SEC for that purpose.

If the EMT takes the form of a perfect fluid \eqref{eq: EMT perfect fluid}, the ECs can be written in terms of the fluid energy density $\rho$ and pressure $p$ as \cite{Martin-Moruno:2017exc,Bouhmadi-Lopez:2019zvz}
\begin{equation}\label{eq: perfect fluid ECs}
    \begin{array}{rcl}
         \text{(i)} & \text{DEC:} & \rho \geq \abs{p} \geq 0\,,\\
        \text{(ii)} & \text{WEC:} &  \rho + p \geq 0 \,,\quad \rho \geq 0\,,\\
        \text{(iii)} & \text{SEC:} & \rho + p \geq 0 \,,\quad \rho + 3p \geq 0 \,,\\
           \text{(iv)} & \text{NEC:} & \rho + p \geq 0 \,.\\
    \end{array}
\end{equation}
As we will see in the next section, this case simplifies the treatment of the theories investigated in the present work.

\subsection{Review of EMT conservation}
\noindent The EMT conservation equations are written as
\begin{equation}\label{eq: EMT conservation}
    \nabla_\alpha T^{\alpha\nu} = 0 \,.
\end{equation}
When working with a perfect fluid, it is common practice to project them onto the directions longitudinal and transverse to the fluid's velocity. For the former, one must simply contract with the velocity $u_\nu$, while for the latter one must act with the orthogonal projector $h\indices{^\mu_\nu} = \delta^\mu_\nu - u^\mu u_\nu$. As a result, we respectively have \cite{hawking_large_2008}
\begin{subequations}
    \begin{equation}\label{eq: EMT conserv sobre u}
        \dot{\rho} + (\rho + p) \nabla_\mu u^\mu = 0 \,,
    \end{equation}
    \vspace{-20pt}
    \begin{equation}\label{eq: EMT conserv sobre h}
        (\rho+p) \dot{u}^\mu - \left( g^{\mu\nu} - u^\mu u^\nu \right) \nabla_\nu p = 0 \,,
    \end{equation}
\end{subequations}
where we use the notation $\dot{\,\,} \equiv u^\mu \nabla_\mu$. For a scalar function $\phi$, it is also the case that $\dot{\phi} = u^\mu \partial_\mu \phi =d\phi/d\tau$, with $\tau$ the parameter of the curve.\\
\indent Since later on it will be of interest to study purely transverse equations, we shall here introduce a triplet of linearly independent transverse vectors $\left(w_1^\mu, w_2^\mu, w_3^\mu\right)$, which we collectively denote as $\Vec{w}^\mu$. They satisfy
\begin{equation}\label{eq: w y u transversos}
    u_\mu \Vec{w}^\mu = 0 \,,
\end{equation}
and we understand that it holds for all three of them. We shall also denote the projection of the derivatives on the transverse directions as
\begin{equation}
    \Vec{\nabla} \equiv \Vec{w}^\mu \nabla_\mu \equiv \left( w_1^\mu \nabla_\mu \,,\,\, w_2^\mu \nabla_\mu \,,\,\, w_3^\mu \nabla_\mu \right) \,.
\end{equation}

\section{The perfect fluid approach}\label{section: The perfect fluid approach}
\noindent We shall in this section show that, under the assumption of the field having a time-like derivative $\partial_\mu \psi$, our model is equivalent to considering a perfect fluid. If we wish to write the EMT \eqref{eq: model EMT} in the perfect fluid form \eqref{eq: EMT perfect fluid}, we must find appropriate correspondences between quantities, and a natural one to begin with seems to be
\begin{equation}\label{eq: corresp velocity}
    u^\alpha \equiv \frac{\partial^\alpha \psi}{\sqrt{(\partial\psi)^2}}
\end{equation}
for the velocity. Note that this correspondence only makes sense if the derivative of the field is a time-like vector, and this shall be our assumption throughout the work. Of course, it is a rather strong assumption that does not hold in many interesting situations (such as a static field). Nevertheless, it is a reasonable assumption in cosmological scenarios.\\
\indent In order to find the energy density, it is useful to recall that $T_{\mu\nu} u^\mu u^\nu = \rho$ for a perfect fluid, and so from our EMT \eqref{eq: model EMT} and definition \eqref{eq: corresp velocity} we obtain
\begin{equation}\label{eq: corresp rho}
    \rho = \frac{2}{\sqrt{g}} \left\{ \frac{1}{2} f_k (\partial\psi)^2 + g \left[ f_v' V - \frac{1}{2} f'_k (\partial\psi)^2 \right] \right\} \,.
\end{equation}
Using all of the above in order to find our last unknown, it follows that
\begin{equation}\label{eq: corresp pressure}
    p = -\frac{2g}{\sqrt{g}} \left[ f_v' V - \frac{1}{2} f'_k (\partial\psi)^2  \right] \,.
\end{equation}
With these correspondences, we may translate our EMT \eqref{eq: model EMT} into that of a perfect fluid as we intended, and study its behavior. Let us note here that $\rho$ and $p$ are both TDiff scalars. We shall now study our two limiting regimes of potential and kinetic domination in the context of the perfect fluid approach.

\subsection{Potential domination in the perfect fluid}
\noindent In the potential domination regime of our perfect fluid, we find that $\rho$ and $p$ are related by a characteristic equation of state (EoS):
\begin{equation}\label{eq: rho and p in potential dominated}
    p = - \rho = -2 \sqrt{g} \, f'_v V \,,
\end{equation}
that is, we have a barotropic fluid $p=p(\rho)$ with the simple EoS $p = w \rho$, where $w = -1$ for any function $f_v$.

\subsection{Kinetic domination in the perfect fluid}
\noindent The kinetic domination regime of our perfect fluid leads to $\rho$ and $p$ taking the simplified form
\begin{equation}\label{eq: rho y p kinetic}
    \rho = \frac{(\partial\psi)^2}{\sqrt{g}} \left( f_k - g f'_k \right) \,, \quad
    p = \frac{(\partial\psi)^2}{\sqrt{g}} \, g f'_k \,.
\end{equation}
It thus follows that we again have a barotropic fluid, whose EoS parameter may now be expressed as
\begin{equation}\label{eq: EoS kinetic}
    w = \frac{p}{\rho} = \frac{g f'_k}{f_k - g f'_k} \,,
\end{equation}
and it is interesting to note that the only dependence is in the metric determinant, $w=w(g)$. Finally, let us also introduce for future reference the variable
\begin{equation}\label{eq: definition of F}
    F \equiv \frac{gf'_k}{f_k} \,,
\end{equation}
in terms of which the EoS parameter in the kinetic regime is written
\begin{equation}\label{eq: EoS, w y F}
    w = \frac{F}{1-F} \,.
\end{equation}
We remark that this expression is the generalization of the one obtained in reference \cite{maroto_tdiff_2023} for the cosmological case. Note that the GR limit (i.e. Diff invariance) implies that $f_k \propto \sqrt{g}$ $\Leftrightarrow$ $F=1/2$ $\Leftrightarrow$ $w=1$. In other words, in the GR limit of the kinetic domination regime we have stiff matter ($p = \rho$). However, in a TDiff theory the coupling function may take on a different form, so one will have a different equation of state for the scalar field. In our study we keep this function arbitrary to show the possible equation of state parameters that one can describe depending on what $f_k$ is chosen. This, in turn, gives rise to a wide range of possibilities, for example, for dark sector models.

\section{Energy conditions}\label{section: Energy conditions}
\noindent In this section we focus on the ECs of our model from the perfect fluid description. Before doing so, however, we remark that with the EMT written in the completely general form \eqref{eq: model EMT}, the NEC is translated into
\begin{equation}\label{eq: NEC general}
    f_k \geq 0 \,.
\end{equation}
We thus find that the NEC is satisfied whenever the kinetic term is non-negative or, in other words, whenever it is not a ghost field \cite{Sbisa:2014pzo}. On another note, as we saw, the network of implications among the ECs (DEC $\Rightarrow$ WEC $\Rightarrow$ NEC $\Leftarrow$ SEC) means that if the NEC is violated then so are all the others, and so the minimum requirement if we wish to satisfy some of the ECs is simply that $f_k \geq 0$.\\
\indent Let us now see what the perfect fluid ECs \eqref{eq: perfect fluid ECs} translate into. The WEC may be written as
\begin{equation}
    f_k \geq 0 \quad \text{\&} \quad \frac{(\partial\psi)^2}{2}\left( f_k - g f'_k \right) + g f'_v V \geq 0 \,,
\end{equation}
while the SEC becomes
\begin{equation}
    f_k \geq 0 \quad \text{\&} \quad \frac{(\partial\psi)^2}{2}\left( f_k + 2g f'_k \right) - 2g f'_v V \geq 0 \,.
\end{equation}
Regarding the DEC, the absolute value of the pressure we find in \eqref{eq: perfect fluid ECs} implies that this condition splits into two possible cases as follows:
\begin{equation}
\begin{split}
    p\leq 0\,:& \quad
        f'_v V - \frac{(\partial \psi)^2}{2} f'_k \geq 0 \quad\&\quad
        f_k \geq 0 \,,\\[5pt]
    p > 0\,: & \quad
        f'_v V - \frac{(\partial \psi)^2}{2} f'_k < 0 \quad\&\quad   \frac{(\partial\psi)^2}{2}\left( f_k - 2g f'_k \right) + 2g f'_v V \geq 0 \,.
\end{split}
\end{equation}
With the ECs written as above, we have full generality. Nevertheless, in order to simplify their treatment and gain some further insight on their physical implications, we shall now study our two limiting cases of potential and kinetic domination.

\subsection{ECs in potential domination}
\noindent We previously found that in the potential domination regime of our perfect fluid the EoS was simply $w=-1$. It is easy to see that for such an EoS the NEC is trivially satisfied (the inequality saturates), while the others translate to:
\begin{equation}
    \begin{gathered}
        \text{WEC:} \quad f'_v V \geq 0 \,,\quad \text{SEC:} \quad f'_v V \leq 0 \,, \\[5pt]
        \text{DEC:} \quad \left\{ \,\,\begin{split}
            p\leq 0 \,:&\quad f_v' V \geq 0 \,, \\
            p > 0 \,:&\quad f_v' V < 0\quad \& \quad f_v' V \geq 0 \quad(\text{contradiction})
    \end{split} \right.
    \end{gathered}
\end{equation}
We note first of all that if the pressure is negative or zero then the DEC is equivalent to the WEC (since the second inequality trivially saturates), whereas if it is positive then the DEC can never be satisfied. Therefore, one can conclude that physically reasonable matter should have non-positive pressure in this case. On the other hand, we also find that the only case where the ECs could be simultaneously satisfied is when they saturate, that is $f'_v V = 0$. This either means that $V = V_0 = 0$ (which cannot happen if we want potential domination) or $f_v(g) =$ const. It is interesting to note that in the latter case the vanishing of the EMT \eqref{eq: EMT in potential dom} means that a scalar field with only a potential term that couples through a constant function does not gravitate (in the sense that it does not affect the geometry of spacetime). Thus, even though it satisfies all ECs, such a matter field would be impossible to be detected through gravitational observations. 

The discussion above implies that, in the potential domination regime, reasonable matter which has nontrivial gravitational effects necessarily violates the SEC. This is not surprising since it is what happens with a vacuum energy, whose EMT takes the form $T_{\mu\nu} \propto g_{\mu\nu}$ as in the present case. Indeed, as we will see in section \ref{section: Conservation of the EMT}, the field equation and the conservation equation imply that the energy density is constant in the potential domination regime.
Therefore, TDiff scalar fields in the potential limit will behave as a cosmological constant.

\subsection{ECs in kinetic domination}
\noindent As we have demonstrated for the general case, the NEC implies $f_k\geq 0$. Nevertheless, $f_k \neq 0$ in kinetic domination, and hence we must consider $f_k>0$. The rest of the ECs translate as:
\begin{equation}\label{eq: kinetic ECs}
    \begin{gathered}
        \text{WEC:} \quad f_k \geq 0 \,,\quad f_k - g f_k' \geq 0\,,\\[5pt]
        \text{SEC:} \quad f_k \geq 0 \,,\quad f_k + 2 g f_k' \geq 0\,,\\[5pt]
        \text{DEC:} \quad \left\{ \,\,\begin{split}
            p\leq 0\,:&\quad f_k' \leq 0 \quad \&\quad f_k \geq 0 \,, \\
            p > 0 \,:&\quad  f_k' > 0 \quad \& \quad f_k - 2g f_k' \geq 0 \,.
    \end{split} \right.
    \end{gathered}
\end{equation}
It is straightforward to check that all ECs considered in this work are indeed satisfied in this regime for the GR case $f_k(g) \propto \sqrt{g}$. This is not surprising since, as we have already mentioned, this case corresponds to stiff matter. (Note, however, that stiff matter violates the abandoned Trace Energy Condition \cite{Barcelo:2002bv}.) On the other hand, as it is not possible to discuss the potential fulfillment of the ECs for general kinetic coupling functions, we will postpone such discussion until section \ref{section: Some particular cases}, where we consider some particular cases of interest with a more immediate interpretation.

\section{Conservation of the EMT}\label{section: Conservation of the EMT}

\noindent In GR the conservation of the EMT on the solutions to the EoM is an immediate consequence of the full Diff invariance of the theory, and hence provides no additional information. In a TDiff theory, where the symmetry is broken, it is therefore a legitimate question to ask what happens to such a conservation: it is not fulfilled automatically, but can it be recovered? And, if so, does it provide any valuable information?

In this work, the gravitational action is unchanged and we consider a single TDiff scalar field as our matter content. As we saw,
\begin{equation}
    S = S_\text{EH} + S_m \,\implies \, G_{\mu\nu} = 8\pi G \, T_{\mu\nu} \,.
\end{equation}
The equations of motion for the gravitational field are the Einstein equations, and since the Bianchi identities establish the divergenceless character of the Einstein tensor, then so must be the field EMT:
\begin{equation}
    \nabla_\mu G^{\mu\nu} = 0 \,\implies\, \nabla_\mu T^{\mu\nu} = 0 \,.
\end{equation}
The field EMT conservation equation on the solutions to the EoM of the theory thus arises as a consistency requirement. It will not be trivially satisfied (we do not have the symmetry to obtain it automatically), but rather it will impose some constraints on the metric tensor.
    
\indent The conclusion seems reasonable. Note that in a Diff invariant theory we have 4 gauge degrees of freedom, which allow us to fix 4 components of the metric via  coordinate transformations. On the other hand, in a TDiff theory, the constraint $\partial_\mu \xi^\mu = 0$ on the allowed transformations implies that we only have 3 gauge degrees of freedom, and hence only 3 components of the metric may be fixed (see references \cite{Alvarez:2007nn,maroto_tdiff_2023} for a discussion in a cosmological context). The additional metric component is actually physical, so we need an extra equation to find it: this is the constraint we obtain from the fact that the EMT conservation equation of our TDiff scalar field is not trivially satisfied. We shall consider in this section our limiting cases of potential and kinetic domination and obtain the corresponding constraints on the metric.

\subsection{EMT conservation in potential domination}
\noindent The conservation of the EMT in potential domination is quite straightforward as a consequence of the simple EoS $w = -1$. Indeed, in this case the longitudinal projection \eqref{eq: EMT conserv sobre u} is simply $\dot{\rho} = 0$, which in turn also means that $\dot{p} = -\dot{\rho} = 0$. Taking this into account, the transverse projection \eqref{eq: EMT conserv sobre h} simplifies down to $\partial^\mu p = 0$. Now substituting equation \eqref{eq: rho and p in potential dominated} in $\partial^\mu p = 0$, and recalling that $V=V_0=$ const. on the solutions to the EoM, we obtain
\begin{equation}\label{eq: potential constraint}
    \left( \frac{1}{2} f'_v + g f_v'' \right) \partial_\mu g = 0 \,.
\end{equation}
In this way, if the coupling function $f_v$ is left arbitrary, we obtain the sought-for constraint on the metric: we must have a constant determinant,
\begin{equation}\label{eq: g const}
    \partial_\mu g = 0 \, \implies \, g = \text{const.}
\end{equation}

Another possibility is to allow the metric determinant to change, but then we must require
\begin{equation}\label{eq: fv in pot}
    f_v(g) = A \sqrt{g} + B \,,
\end{equation}
with $A$ and $B$ constants of integration. Thus, only particular theories (with this $f_v$) do not restrict the form of $g$ due to the conservation of the EMT. Expression \eqref{eq: fv in pot} is the most general solution for this situation, and different limiting cases may be explored. On the one hand, if we set $B=0$, we recover GR as a particular solution ($f_v \propto \sqrt{g}$) as it should be expected. On the other hand, setting $A=0$ means that the coupling is done via a constant function which, as we previously discussed, leads to a vanishing EMT and therefore the field does not gravitate.\\
\indent As a final comment, we note that no further information may be gained from the longitudinal projection of the EMT conservation, since substituting \eqref{eq: rho and p in potential dominated} in $\dot{\rho} = 0$ yields
\begin{equation}
    \left( \frac{1}{2} f'_v + g f_v'' \right) \frac{dg}{d\tau} = 0 \,,
\end{equation}
which is trivially satisfied using equation \eqref{eq: potential constraint}.

\subsection{EMT conservation in kinetic domination}
\noindent The study of the EMT conservation in the kinetic domination regime turns out to be rather more involved, and for this reason we have divided the analysis into smaller, more accessible parts. Firstly, we will rewrite the kinetic EoM in terms of perfect fluid quantities, and also find what its solution must satisfy. We will also consider the longitudinal and transverse projections of the EMT conservation equation for our particular case, and separately study them in order to see what constraints the EMT conservation imposes on the theory. In passing, we will obtain a very simple expression for the energy density and the speed of sound of the perturbations.

\subsubsection{EoM and conservation equations}
\noindent In this introductory section we present some results regarding the EoM, as well as the longitudinal and transverse projections of the EMT conservation equation. The  subsequent sections shall be devoted to a more thorough study of the results we find here, this one simply serves as their presentation.

We shall begin by expanding the kinetic EoM \eqref{eq: EoM kinetic} as
\begin{equation}\label{eq: expanded EoM}
    0 = \partial^\mu\psi \partial_\mu f_k + f_k \partial_\mu\partial^\mu \psi \,.
\end{equation}
We shall denote the normalization constant of the velocity by $N = \sqrt{(\partial\psi)^2}$, so that $\partial^\alpha \psi = N u^\alpha$, and divide the above equation by $f_k$. For now, we shall assume $f_k \neq 0$ everywhere until the end of the section, where we will discuss the possibility of the function vanishing at some point. So, we obtain
\begin{equation}\label{eq: expanded kinetic EoM}
    0 = N u^\mu\partial_\mu \left(\ln \abs{f_k} \right) + \partial_\mu\left(N u^\mu \right) \,.
\end{equation}
Using equation \eqref{eq: divergence nabla and partial}, we can write the second term as
\begin{equation}
    \partial_\mu \left( \frac{N u^\mu \sqrt{g}}{\sqrt{g}}  \right)  = N\left( u^\mu \partial_\mu \ln\frac{N}{\sqrt{g}} + \nabla_\mu u^\mu \right) \,.
\end{equation}
In this expression we recognize the expansion scalar of the congruence, $\theta = \nabla_\mu u^\mu$. This quantity may be directly related to the fractional rate of change of the congruence's cross-sectional volume $\delta V$ as \cite{poisson_relativist39s_2004}
\begin{equation}\label{eq: cross-sectional volume}
    \theta = \frac{1}{\delta V} \frac{d}{d\tau}(\delta V) = u^\mu\partial_\mu \ln \delta V \,.
\end{equation}
Taking everything into account, we may rewrite equation \eqref{eq: expanded kinetic EoM} (after simplifying a common factor of $N$) as
\begin{equation}\label{eq: EoM reescritas para usar luego}
    \frac{d}{d\tau} \ln \abs{\frac{f_k}{\sqrt{g}} N \delta V } = 0 \,,
\end{equation}
where we have used $u^\mu\partial_\mu = \frac{d}{d\tau}$. Recalling $N=\sqrt{(\partial\psi)^2}$, it finally follows that the solutions to the EoM satisfy
\begin{equation}\label{eq: solutions to EoM}
    (\partial \psi)^2 = C_\psi(x) \left( \frac{\sqrt{g}}{\delta V f_k}\right)^2 \,,
\end{equation}
with $C_\psi(x)$ a function subject to the constraint
\begin{equation}
    \dot{C}_\psi(x) = u^\mu\partial_\mu C_\psi(x) = 0 \,.
\end{equation}
\indent It will also be useful to rewrite the EoM in terms of perfect fluid quantities, and combine it with the longitudinal projection of the EMT conservation equation. Taking into account equations \eqref{eq: rho y p kinetic}, we may express $N$ in terms of $\rho$ and $p$ as
\begin{equation}\label{eq: N i.t.o. rho and p}
    N^2 = \frac{(\rho+p)\sqrt{g}}{f_k}  \,,
\end{equation}
and so the EoM \eqref{eq: EoM reescritas para usar luego} takes the form
\begin{equation}
    \frac{d}{d\tau} \ln  \sqrt{\frac{f_k}{\sqrt{g}} (\rho + p) \delta V^2}  = 0 \,.
\end{equation}
From this expression we find that
\begin{equation}\label{eq: dot rho + dot p}
    \dot{\rho} + \dot{p} = - (\rho + p) \, \frac{d}{d\tau} \ln  \abs{\frac{\delta V^2 f_k}{\sqrt{g}} } \,.
\end{equation}
But we also know, from equation \eqref{eq: EMT conserv sobre u}, that
\begin{equation}\label{eq: dot rho longitudinal}
    \dot{\rho} = -(\rho+p) \, \frac{d}{d\tau} \ln \delta V \,,
\end{equation}
and so from equations \eqref{eq: dot rho + dot p} and \eqref{eq: dot rho longitudinal}, it follows:
\begin{equation}\label{eq: dot p longitudinal}
    \dot{p} = - (\rho + p) \, \frac{d}{d\tau} \ln \abs{ \frac{\delta V f_k}{\sqrt{g}}} \,.
\end{equation}
Up to this point, we have been dealing with longitudinal results, in the sense that they hold along the integral curves of the tangent vector field $u^\mu$. So far, however, we have no information as to what happens in directions transverse to the fluid's velocity, so let us consider precisely that.\\
\indent We begin with equation \eqref{eq: EMT conserv sobre h}, the projection of the EMT conservation equation onto directions transverse to the fluid's velocity. In that expression, we recognize terms with $\dot{u}^\mu$, $u^\mu$, and $\partial^\mu p$, but it turns out that these three quantities are related.  It follows from the definition of the pressure in equation \eqref{eq: rho y p kinetic} that
\begin{equation}\label{eq: rewrite pressure}
    p = N^2 \frac{g f_k'}{\sqrt{g}} = \frac{ N^2 F f_k}{\sqrt{g}} \,,
\end{equation}
where $F$ was defined in equation \eqref{eq: definition of F}. Since we wish to take its derivative, let us begin by computing the following:
\begin{equation}\label{eq: deriv de N^2}
    \begin{split}
        \partial^\mu (N^2) &= \nabla^\mu (N^2)= \nabla^\mu (\nabla_\alpha \psi \nabla^\alpha \psi) = 2 \nabla_\alpha \psi \nabla^\mu \nabla^\alpha \psi =\\[5pt]
        &= 2 N u_\alpha \nabla^\alpha (N u^\mu) = 2N ( \dot{N} u^\mu + N \dot{u}^\mu) =\\[5pt]
        &=  2N^2 \left( u^\mu \frac{d}{d\tau}\ln N + \dot{u}^\mu \right) \,,
    \end{split}
\end{equation}
where we have recalled that for a torsionless connection $\nabla^\mu \nabla^\alpha \psi = \nabla^\alpha \nabla^\mu \psi$ when acting on scalar functions. Using this result when differentiating equation \eqref{eq: rewrite pressure}, we find
\begin{equation}
    \partial^\mu p = 2p \left( u^\mu \frac{d}{d\tau}\ln N + \dot{u}^\mu + \frac{1}{2}\partial^\mu \ln \abs{\frac{F f_k}{\sqrt{g}} } \right) \,.
\end{equation}
From this expression we may solve for $\dot{u}^\mu$,
\begin{equation}\label{eq: u dot}
    \dot{u}^\mu = \frac{1}{2} \partial^\mu \ln \abs{\frac{p \sqrt{g}}{F f_k} } - u^\mu \frac{1}{2} \frac{d}{d\tau} \ln N^2 \,.
\end{equation}
Now, using equations \eqref{eq: N i.t.o. rho and p} and \eqref{eq: rewrite pressure} we can write
\begin{equation}
    N^2 = \frac{p \sqrt{g}}{F f_k} = \frac{(\rho+p) \sqrt{g}}{f_k} \,,
\end{equation}
and inserting the resulting $\dot{u}^\mu$ in \eqref{eq: EMT conserv sobre h}, we obtain the transverse equation
\begin{equation}\label{eq: transverse equation forreal}
    0 = u^\mu \dot{p} - \partial^\mu p  + \frac{\rho+p}{2}\left( \partial^\mu \ln \frac{(\rho+p) \sqrt{g}}{f_k} -u^\mu \frac{d}{d\tau} \ln \frac{(\rho+p) \sqrt{g}}{f_k} \right) \,.
\end{equation}
It is immediate to see that the equation is trivially satisfied if we contract with $u_\mu$, as we should expect from transversality.\\
\indent Let us now assume that $f_k$ vanishes at some point $\mathcal{P}$ in spacetime, $f_k |_\mathcal{P} =0$. Assuming a regular $f_k'$, the EoS \eqref{eq: EoS kinetic} read at that point implies $w |_\mathcal{P} =-1$. The field EoM \eqref{eq: expanded EoM} \textit{at that point} are written
\begin{equation}
     0 = \partial^\mu\psi \partial_\mu f_k = f'_k \partial^\mu\psi \partial_\mu g \,\implies\, \dot{g}|_\mathcal{P} = 0 \,.
\end{equation}
On the other hand, the conservation of the energy \eqref{eq: EMT conserv sobre u} reads $\dot{\rho}|_\mathcal{P} = 0$ due to the simple EoS at that point. Taking now into account the definition of the energy density in the kinetic regime \eqref{eq: rho y p kinetic}, and evaluating its variation along the curve at that point, it follows that
\begin{equation}
    \left.\frac{d}{d\tau}(\partial\psi)^2\right|_\mathcal{P} = 0 \,,
\end{equation}
which in turn means $\dot{p}|_\mathcal{P} = 0$ due to the definition \eqref{eq: rho y p kinetic}. Therefore, most importantly, we end up having $\dot{w}|_\mathcal{P} = 0$. This means that, at a neighborhood along the curve, $w=-1$ as well. As for the transverse condition, equation \eqref{eq: EMT conserv sobre h} gives $\partial^\mu p|_\mathcal{P} = 0$.\\
\indent On the other hand, let us finally consider the case $f_k \propto g$, so that we have $\rho=0$. Taking into account the conservation equation \eqref{eq: EMT conserv sobre u}, it follows that the pressure also vanishes. In what follows, we shall not consider further this trivial case.

\subsubsection{Longitudinal results}
\noindent It is possible to find a longitudinal constraint on the metric by considering our longitudinal equations \eqref{eq: dot rho longitudinal} and \eqref{eq: dot p longitudinal}, together with the EoS in the kinetic limit \eqref{eq: EoS, w y F}. Using the EoS and equation \eqref{eq: dot rho longitudinal}, the derivative of the pressure is
\begin{equation}
    \dot{p} = \dot{w} \rho + w \dot{\rho} = \dot{w} \rho - w\rho(1+w) \, \frac{d}{d\tau} \ln \delta V \,.
\end{equation}
Equating this expression with equation \eqref{eq: dot p longitudinal}, dividing through by $\rho$, and reorganizing terms we find
\begin{equation}\label{eq: w dot chi dot y delta V}
    \dot{w} = (1+w) \left[ (w-1) \, \frac{d}{d\tau} \ln \delta V - \frac{d}{d\tau} \ln\abs{ \frac{f_k}{\sqrt{g}}} \right] \,.
\end{equation}
Recalling what quantities are functions of only the determinant, it is possible to write
\begin{gather}
\dot{w} = w' \, \dot{g} = (1+w) \frac{F'}{1-F} \, \dot{g}\,, \\[5pt]
\frac{d}{d\tau}\left(\frac{f_k}{\sqrt{g}}\right) = \frac{f_k}{\sqrt{g}} \left( F-\frac{1}{2} \right) \frac{\dot{g}}{g}  \,,
\end{gather}
where we have used equation \eqref{eq: EoS, w y F}. Substitution gives
\begin{equation}
    (1+w) \frac{F'}{1-F} \, \dot{g} = (1+w) \left[ (w-1) \frac{d}{d\tau} \ln \delta V - \left( F-\frac{1}{2} \right) \frac{\dot{g}}{g} \right] \,.
\end{equation}
Dividing through by $(1+w)$ and rearranging, we obtain
\begin{equation}\label{eq: almost finished}
    \left\{ \frac{F'}{1-F} + \frac{F-1/2}{g} \right\} \, \dot{g}= (w-1) \, \frac{d}{d\tau}\ln \delta V \,.
\end{equation}
After dividing this equation by $(w-1)$ ($w=1$ would correspond to considering the GR limit) and taking into account that from equation \eqref{eq: EoS, w y F} we have
\begin{equation}\label{eq: w-1}
    w-1 = \frac{2F - 1}{1-F} \,,
\end{equation}
it turns out that everything simplifies neatly, since
\begin{equation}\label{eq: simplifies neatly}
    \frac{1}{w-1} \left\{ \frac{F'}{1-F} + \frac{F-1/2}{g} \right\} = \frac{1}{2}\left\{ \frac{2F'}{2F-1} + \frac{1}{g} - \frac{f'_k}{f_k} \right\} = \frac{1}{2} \frac{d}{dg} \ln \abs{(2F-1)\frac{g}{f_k}} \,.
\end{equation}
In this way, we have from \eqref{eq: almost finished} that
\begin{equation}
    \frac{\dot{g}}{2} \, \frac{d}{dg} \ln \abs{(2F-1)\frac{g}{f_k}} =\frac{d}{d\tau} \ln \delta V \,.
\end{equation}
Recognizing a total derivative in the LHS and multiplying the whole equation by 2 it follows that
\begin{equation}
    \frac{d}{d\tau} \ln \abs{(2F-1)\frac{g}{f_k}} =\frac{d}{d\tau} \ln \delta V^2  \,.
\end{equation}
Finally, then, we obtain the following longitudinal constraint:
\begin{equation}\label{eq: longitudinal constraint}
    (2F-1)\frac{g}{f_k} = C_g(x) \delta V^2 \,,
\end{equation}
with $C_g(x)$ a function which must satisfy
\begin{equation}
    \dot{C}_g(x) = u^\mu\partial_\mu C_g(x) = 0 \,.
\end{equation}
We emphasize that equation \eqref{eq: longitudinal constraint} imposes a constraint on the metric once a coupling function $f_k$ is specified. Note that the expression obtained generalizes for an arbitrary metric the result obtained in reference \cite{maroto_tdiff_2023} for Robertson-Walker, where $\delta V = a^3$ and $C_g =$ const.

\subsubsection{Transverse results}
\noindent We now focus on the transverse part of the EMT conservation, and in order to do so we contract the transverse equation \eqref{eq: transverse equation forreal} with the previously introduced triplet of transverse vectors $\Vec{w}_\mu$ presented in \eqref{eq: w y u transversos}, obtaining
\begin{equation}
    0 = \frac{\rho+p}{2}\, \Vec{\nabla} \ln \abs{\frac{(\rho+p) \sqrt{g}}{f_k}} - \Vec{\nabla} p
\end{equation}
(we remark that throughout this section $\Vec{\nabla}=\Vec{w}^\mu\partial_\mu$). Simple algebraic manipulations now lead to
\begin{equation}
    \Vec{\nabla}(\rho-p) = (\rho+p) \, \Vec{\nabla} \ln \abs{\frac{f_k}{\sqrt{g}}} \,,
\end{equation}
which using the EoS may as well be written as
\begin{equation}\label{eq: rewrite transverse using EoS}
    (1-w)\Vec{\nabla} \rho - \rho \Vec{\nabla} w = \rho(1+w) \, \Vec{\nabla} \ln \abs{\frac{f_k}{\sqrt{g}}} \,.
\end{equation}
Recalling once again that the only dependence of the EoS parameter $w$ and the ratio $f_k/\sqrt{g}$ is on the metric determinant, we have
\begin{gather}
    \Vec{\nabla} w = w' \, \Vec{\nabla} g = (1+w) \frac{F'}{1-F} \Vec{\nabla} g \,, \\[5pt]
    \Vec{\nabla} \left(\frac{f_k}{\sqrt{g}}\right) = \frac{f_k}{\sqrt{g}} \left( F-\frac{1}{2} \right) \frac{\Vec{\nabla} g}{g} \,,
\end{gather}
where we have again used equation \eqref{eq: EoS, w y F}. Substituting back in equation \eqref{eq: rewrite transverse using EoS} and rearranging, it follows that
\begin{equation}
    (1-w)\Vec{\nabla} \rho = \rho(1+w) \left\{ \frac{F'}{1-F} + \frac{F-1/2}{g} \right\} \Vec{\nabla} g \,.
\end{equation}
We may now safely divide both sides of the equation by $\rho(1-w)$, finally arriving at
\begin{equation}\label{eq: starting point for transverse}
    \Vec{\nabla} \ln\abs{\rho} = -\frac{1+w}{2} \, \Vec{\nabla} \ln \abs{(2F-1)\frac{g}{f_k}}\,.
\end{equation}
where we have also used equation \eqref{eq: simplifies neatly}.\\
\indent Now, we have already obtained a (longitudinal) constraint in equation \eqref{eq: longitudinal constraint}, and we wonder if the transverse projection of the EMT conservation equations might provide us with any additional information. To this end, we shall substitute the longitudinal constraint \eqref{eq: longitudinal constraint} in the RHS of equation \eqref{eq: starting point for transverse}. Recalling also that $1+w = \frac{1}{1-F}$ and that on the solutions \eqref{eq: solutions to EoM} to the EoM
\begin{equation}
    \rho = \frac{(\partial\psi)^2}{\sqrt{g}} f_k(1-F) = \frac{C_\psi \sqrt{g} (1-F)}{f_k \delta V^2} \,,
\end{equation}
it follows that
\begin{equation}\label{eq: starting point for transverse constraint}
    \Vec{\nabla}\ln\abs{\frac{C_\psi \sqrt{g} (1-F)}{f_k \delta V^2}} = \frac{-1}{2(1-F)} \Vec{\nabla} \ln \abs{C_g \delta V^2}\,.
\end{equation}
Working from this expression, a straightforward calculation (see Appendix \ref{section: Calculation of the transverse constraint}) finally yields the transverse constraint
\begin{equation}\label{eq: ligadura transversa}
    \Vec{\nabla}\left( C_g C_\psi \right) = \Vec{w}^\mu\partial_\mu \left( C_g C_\psi \right)= 0 \,.
\end{equation}
This is a condition which relates the function $C_g$ from the longitudinal constraint \eqref{eq: longitudinal constraint} to the solutions of the EoM \eqref{eq: solutions to EoM}, which depend on $C_\psi$. Indeed, we find in \eqref{eq: ligadura transversa} that the derivative of the product $C_g C_\psi$ vanishes when projected onto the transverse directions. However, we know that the derivative of the product also vanishes when projected along the longitudinal direction, since $\dot{C}_\psi = \dot{C}_g = 0$ and hence $u^\mu\partial_\mu\left( C_g C_\psi \right) = 0$. Consequently, we find that
\begin{equation}\label{eq: Cg Cpsi = crho}
    C_g C_\psi = \text{const.} \equiv c_\rho \,,
\end{equation}
i.e. the product is actually a constant (which we have denoted as $c_\rho$ for later convenience) and the two functions are inversely proportional to each other.

As a final note, let us remark for clarity that even though we employ the terms ``longitudinal constraint'' and ``transverse constraint'' in order to more easily refer to said expressions, only one actual condition on the metric has been obtained (as it should) which is equation \eqref{eq: longitudinal constraint}. Equation \eqref{eq: ligadura transversa} simply relates the (in principle rather general) functions which come from integration.

\subsubsection{The adiabatic TDiff fluid}
\noindent We will now derive a simple expression for the energy density $\rho$ in the kinetic domination regime. In order to do so, we begin with its definition in equation \eqref{eq: rho y p kinetic}, substitute the solutions to the EoM \eqref{eq: solutions to EoM}, recall \eqref{eq: w-1}, and use the longitudinal constraint \eqref{eq: longitudinal constraint}:
\begin{equation}
    \rho = \frac{(\partial\psi)^2}{\sqrt{g}} f_k(1-F) = \frac{C_\psi}{\sqrt{g}} \frac{g}{f_k \delta V^2} (1-F) = \frac{C_\psi}{\sqrt{g}} \frac{g}{f_k \delta V^2} \frac{2F-1}{w-1} = \frac{C_g C_\psi}{(w-1)\sqrt{g}} \,.
\end{equation}
Thus, recalling the consequence \eqref{eq: Cg Cpsi = crho} of the transverse constraint, we finally obtain the following simple relation:
\begin{equation}\label{eq: resultado rho}
    \rho = \frac{c_\rho}{(w-1) \sqrt{g}} \,.
\end{equation}
We remark that this expression is well-defined as long as we are not in the GR limit, but other than that it is completely general and valid for all geometries. Moreover, since $w=w(g)$, it also shows that the only dependence of both the energy density and the pressure is in the metric determinant, and this means that the possible perturbations of this fluid shall be adiabatic.

Furthermore, for an adiabatic fluid, the speed of sound of the perturbations $c_s$ is defined through
\begin{equation}
    \delta p = c_s^2 \, \delta\rho \,,
\end{equation}
where it is possible to find that
\begin{equation}
    c_s^2 = w + w' \frac{\rho}{\rho'} \,.
\end{equation}
Substituting, it finally follows that for our case:
\begin{equation}\label{eq: cs^2}
    c_s^2 = -\frac{g f_k (f'_k + 2 g f_k'')}{f_k^2 + (2 g f_k')^2 - gf_k (5 f_k' + 2 g f_k'')} \,.
\end{equation}
Let us emphasize that the above relation for the speed of sound of the perturbations is valid for any scalar field model in the kinetic regime, as well as for any spacetime. In particular, it is useful in the study of cosmological perturbations in TDiff scenarios, and it simplifies the treatment that was made in reference \cite{maroto_tdiff_2023}. We remind the reader that the only underlying assumption is that the field derivative is a time-like vector.

\section{Shift-symmetric TDiff dark sector}\label{section: Some particular cases}
In this section we focus our attention on a particular subclass of models in which the kinetic domination is not an approximate regime but rather an exact behavior. These models are invariant under shift redefinitions of the field
\begin{equation}
    \psi \rightarrow\psi+c,
\end{equation}
where $c$ is an arbitrary constant. Assuming this symmetry is maintained in the quantum regime, the absence of a potential term is protected against radiative corrections. Moreover, these models are of particular interest in different physical situations.
In a cosmological context they produce kinetically driven cosmic dynamics \cite{Germani:2016gzh,Germani:2017pwt}, thus avoiding self-tuning problems coming from relying in a particular form of a potential. Furthermore, shift-symmetric scalar-field theories have also been investigated in detail in black hole scenarios \cite{Ogawa:2015pea,Benkel:2016rlz}. We shall in what follows consider some particular cases of physical interest.

\subsection{Constant equation of state models}
We begin by considering that the coupling function has the form of a power-law, i.e.
\begin{equation}
    f_k(g) = C g^\alpha \,,
\end{equation}
with $C$ and $\alpha$ some constant parameters. Note that, in such a case, $F=gf_k'/f_k=\alpha$ and so it follows that the EoS parameter is also constant:
\begin{equation}
    w = \frac{\alpha}{1-\alpha} \,.
\end{equation}
The case $\alpha = 1$ (which corresponds to $f_k \propto g$) implies $\rho=0$ and was previously discussed.
Away from the GR limit, we find from equation \eqref{eq: resultado rho} that the energy density in power-law couplings satisfies
\begin{equation}
    \rho \propto \frac{1}{\sqrt{g}} \,,
\end{equation}
while the longitudinal constraint \eqref{eq: longitudinal constraint} gives
\begin{equation}
    g \propto (C_g \delta V^2)^\frac{1}{1-\alpha} = (C_g \delta V^2)^{(1+w)}\,.
\end{equation}
We note that if $C_g(x) =$ const. the situation simplifies:
\begin{equation}
    g \propto \delta V^{2(1+w)} \,\implies\, \rho \propto \delta V^{-(1+w)} \,,
\end{equation}
and this may be of use in cosmological settings.

\begin{figure}
    \centering
    \includegraphics[width=0.55\textwidth]{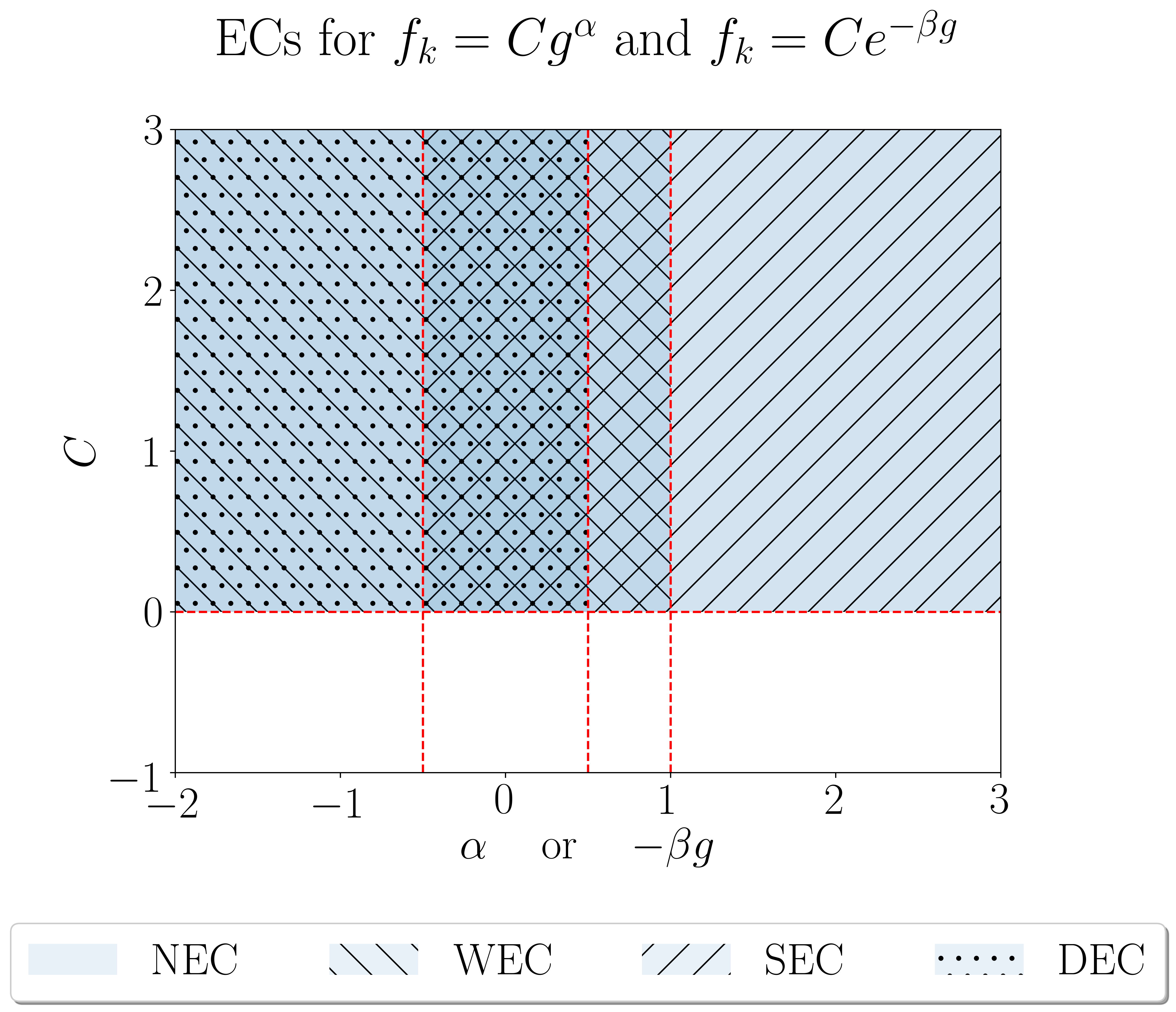}
    \caption{Regions of validity of the ECs for the two couplings (the axes extend infinitely). The same diagram is valid for both, understanding the horizontal axis as $\alpha$ when considering the power-law and as $-\beta g$ when considering the exponential.}
    \label{fig: ECs}
\end{figure}

Next we focus on the ECs \eqref{eq: kinetic ECs}. Noting that we need $C\neq 0$ in order to have a non-vanishing $f_k$, the ECs translate to
\begin{equation}
    \begin{split}
        \text{NEC:} & \quad C > 0\,,\\
        \text{WEC:} & \quad C > 0 \,,\quad \alpha \leq 1 \,,\\
        \text{SEC:} & \quad C > 0 \,,\quad \alpha \geq -1/2\,,\\
        \text{DEC:} & \quad C > 0 \,,\quad \alpha \leq 1/2\,.
    \end{split}
\end{equation}
They are graphically represented in figure \ref{fig: ECs}. For $C>0$ and $\alpha<-1/2$ all the ECs are satisfied except for the SEC. So, we will have non-negative energy densities propagating in a causal way (as seen by any observer), but not necessarily leading to the focusing of time-like geodesics. The corresponding couplings, therefore, could appear interesting for describing dark energy models when applied to a cosmic background. However, when reflecting about this possibility one may quickly note that, since $w$ is constant, one obtains that the propagation speed of the field perturbations is
\begin{equation}
    c_s^2=w \,,
\end{equation}
as it could be obtained from the general relation \eqref{eq: cs^2}. So, stressing that the field perturbations are adiabatic, one can conclude that these dark energy models will be unstable. 
On the other hand, for $C > 0$ and $\alpha \in [-1/2,1/2]$ all of the ECs are satisfied. The particular case of $\alpha=1/4$ may be of interest for the dark sector, as being able describe dark radiation ($w=1/3$) \cite{Bernal:2016gxb}. 
Finally, we note that a similar analysis as that followed for the power-law model could also apply to a more general coupling function which may be expressed as a power series. 

\subsection{Dark matter}
Let us now take a constant coupling function, that is
\begin{equation}
    f_k=C.
\end{equation}
This corresponds to the case $\alpha = 0$ of the previous section. For this model equations \eqref{eq: rho y p kinetic} reduce to
\begin{equation}\label{rhoDM}
    \rho=C\frac{(\partial\psi)^2}{\sqrt{g}} \,,\quad p=0\,,
\end{equation}
while we also have $w=0$. Therefore, in any situation that the scalar field behaves as a perfect fluid (that is, whenever $\partial_\mu\psi$ is time-like) this theory describes dust matter, and as such it is a very simple model for dark matter which does not add new parameters to a cosmological model with respect to $\Lambda$CDM. Moreover, since this is an adiabatic fluid with constant equation of state parameter $w=0$, then the speed of sound vanishes not only in cosmological backgrounds \cite{maroto_tdiff_2023} but in general, this is,
\begin{equation}
    c_s^2 = 0 \,.
\end{equation}
Finally, the simple equation for the energy density \eqref{eq: resultado rho} in this case takes the form
\begin{equation}
    \rho = - \frac{c_\rho}{\sqrt{g}}.
\end{equation}
The compatibility between these equations requires
\begin{equation}
    (\partial\psi)^2=-\frac{c_\rho}{C} \,.
\end{equation}
Since we are assuming the field derivative to be a time-like vector, i.e. $(\partial\psi)^2 > 0$, this means that the constants $c_\rho$ and $C$ have opposite signs.

\subsection{Models with different gravitational domains}
\noindent We now consider that the coupling function is an exponential. This is
\begin{equation}
    f_k(g) = C e^{-\beta g} \,,
\end{equation}
with $C$ and $\beta$ some constant parameters. In this case, the variable $F = -\beta g$ is not a constant, and so neither is the EoS parameter $w$, which from equation \eqref{eq: EoS, w y F} reads
\begin{equation}
    w = - \frac{\beta g}{1 + \beta g} \,.
\end{equation}
Taking into account the general expression for the speed of sound \eqref{eq: cs^2}, we also find that
\begin{equation}
    c_s^2 = \frac{\beta g (1-2\beta g)}{1 + 5\beta g + 2 \beta^2 g^2} \,.
\end{equation}
The study of the evolution of the energy density and the metric determinant is not particularly simple or illuminating (except perhaps for the case $\beta=0$, which gives a constant $f_k$ and non-relativistic matter, as was discussed in the previous section), so we focus on the ECs \eqref{eq: kinetic ECs}. These take the form:
\begin{equation}
    \begin{split}
        \text{NEC:} & \quad C > 0\,,\\
        \text{WEC:} & \quad C > 0 \,,\quad -\beta g \leq 1 \,,\\
        \text{SEC:} & \quad C > 0 \,,\quad -\beta g \geq -1/2\,,\\
        \text{DEC:} & \quad C > 0 \,,\quad -\beta g \leq 1/2\,,
    \end{split}
\end{equation}
where, as in previous cases, we have taken $C\neq 0$ to avoid $f_k=0$. We may graphically represent them in the same way as we did the power-law, see figure \ref{fig: ECs}. Nevertheless, the fact that the metric determinant explicitly appears in these ECs means that the fulfillment of the ECs will depend on the value of $g$, as opposed to the previous case. Since the metric determinant is allowed to vary, it may happen that in some regions of the spacetime a given EC is satisfied but at others it is not. 

As a physically interesting example, let us suppose that $\beta >0$, so that the product $-\beta g$ is always negative. In this case, the WEC and the DEC are always satisfied; so, we have some physically reasonable kind of perfect fluid. However, there may be some regions in spacetime where the SEC holds, whereas in other regions the SEC would be violated. Noting that the fulfillment of the SEC implies focusing of time-like geodesics through the Raychaudhuri equation \cite{Kar:2006ms}, one could find different gravitational domains in spacetime, characterized by focusing or possible defocusing of time-like geodesics. In a cosmological context the transition between those regions would take place at a particular cosmic time; therefore, one may be able to describe cosmological models evolving from a phase of decelerated expansion to other of accelerated expansion \cite{maroto_tdiff_2023}. This is of particular interest for the construction of unified dark matter-energy models \cite{david}.

\subsection{Beyond the shift-symmetric case}\label{sec: beyond}
Let us now go beyond the shift-symmetric case previously discussed. Throughout the paper we have presented different interesting results for general models in the potential domination regime. In the first place, in section \ref{section: Preliminary concepts} we have discussed that the field equation implies $V=V_0$. In the second place, the conservation of the EMT on the solutions to the field equation implies a constant determinant for generic $f_v$ functions. Summarizing, for generic TDiff scalar theories the fluid is such that
\begin{equation}
    p = - \rho = \frac{-2g}{\sqrt{g}} \, f'_v V ={\rm constant}\,,\quad \text{in the potential regime}.
\end{equation}
In this way, one can conclude that these models behave as a cosmological constant in the potential domination regime. 

On the other hand, it is difficult to extract conclusions regarding the gravitational properties of general models in the kinetic regime. Nevertheless, the exact results presented in the previous sections for shift-symmetric models can be regarded as the approximate behavior of general models when approaching the kinetic domination regime. Therefore, one could in principle think that general models may be able to interpolate between the kinetic and the potential limits. In a cosmological context, the presented study can allow one to investigate the possible construction of unified dark matter-energy models when considering $f_v\neq0$ and $f_k=C$, for example.

\section{Discussion and conclusions} \label{section: Discussion and conclusions}
In this work, we have explored the consequences of breaking the Diff invariance of the matter sector down to TDiff in general contexts (i.e. the analysis has been purely geometrical, without assuming a spacetime metric). We have considered a scalar field model which couples to gravity via arbitrary functions of the metric determinant, and studied its limiting cases of potential and kinetic domination. Under the assumption of the field derivative being a time-like vector, we have also shown that it is possible to carry out an equivalent description of the model in terms of a perfect fluid. The ECs have also been analyzed in each of the two regimes, and we have obtained expressions involving derivatives of the coupling functions and their relations to quantities such as the metric determinant (in the case of $f_k$) or the potential (in the case of $f_v$). 

An important focus of the work has been the study of the EMT conservation, which is not automatically satisfied in a TDiff theory, and we have found that it implies a constraint on the metric. For the potential domination regime we have obtained that either the metric determinant has to be constant, or the coupling function has a particular form (which contains GR as a particular limit) to allow the metric determinant to vary.  The study of EMT conservation in the kinetic domination regime resulted in not only the sought-for constraint on the metric tensor (which in this regime turns out to depend on the coupling function chosen), but it also provided us with a particularly simple expression for the energy density (which is not valid in GR) in terms of the metric determinant, which could be very useful in perturbative analysis. Indeed, one of the main results of the work is the fact that in the kinetic regime the energy density and pressure depend only on the metric determinant and, therefore, possible perturbations of the fluid will be adiabatic as a result of this dependence on a single variable. In fact, we have also found a general expression for the speed of sound in terms of the coupling function valid for any geometry.

Let us remark again the fact that the couplings to gravity which we have considered are minimal. Furthermore, the kinetic term is a canonical one, but in situations in which it dominates we have a rather flexible EoS thanks to the variety of coupling functions $f_k$. In GR the only possibility for a shift-symmetric scalar field with a canonical kinetic term is $w=1$, which means that one must include non-canonical kinetic terms in order to allow for interesting phenomenology (this, is for instance, the case of $k$-inflation \cite{Armendariz-Picon:1999hyi} or $k$-essence \cite{amendola_dark_nodate,Armendariz-Picon:2000ulo}, and kinetic gravity braiding \cite{Deffayet:2010qz,BorislavovVasilev:2022gpp}). 

We have shown that shift-symmetric TDiff scalar models with a power-law coupling function are of special interest, since the equation of state parameter is constant in this scenario. For these models, the energy density is simply inversely proportional to the square root of the metric determinant. Regarding the dark sector, this kind of coupling can be applied to describe dark radiation. Furthermore, in the particular case that the coupling function is simply a constant, the model could describe dark matter in generic backgrounds.
On the other hand, we have also considered an exponential coupling function for shift-symmetric TDiff scalar models. In this case the study of the ECs reveals that the evolution of the metric determinant gives rise to particular models that can cross from regions where some EC is satisfied to other where it can be violated. In particular, we conclude that exponential models with a negative exponent satisfy all energy conditions except for the SEC, which will be violated or satisfied in different regions depending on the value of the metric determinant. Therefore, the spacetime could be divided in gravitational domains, depending on the focusing or possible defocusing of time-like geodesics. This is particularly interesting in cosmological contexts, where one could describe the transition from a non-accelerated expansion to an accelerated one. So, these models may be able to provide us with a fundamental description for unified dark matter-energy models \cite{david}.

It is difficult to extract general consequences beyond the shift-symmetric case. For this purpose it is particularly interesting to reflect about the results we have obtained in the potential domination regime and in the kinetic domination regime (which correspond to those obtained for the shift-symmetric models). Therefore, for example, general models with a power-law kinetic coupling may be able to interpolate between a cosmological constant behavior and that of a component with a constant equation of state parameter, being of particular interest the case of dark matter.

Before concluding, let us shortly perform some further comments on the model and future outlook. We begin by noting that, for simplicity, we have only considered a single TDiff scalar field as our matter content, since our study in this work centers around the possibility of describing the dark sector from a novel point of view (in particular, modeling it via a scalar field with symmetry-breaking couplings). Of course, one could well argue that we are missing all the visible matter from the Standard Model in our treatment, and that we should include $S_\text{SM}$ in the total action. Nevertheless, if we perform this inclusion in the standard way, then the newly included piece $S_\text{SM}$ would be Diff invariant all on its own, meaning that the associated EMT is conserved all on its own. This fact tells us that the consistency requirement of the TDiff scalar field $\psi$'s EMT conservation would still follow from the Einstein equations together with the Bianchi identities, and that we did not lose generality in our discussion. How about the inclusion of other TDiff fields of arbitrary character (other scalar fields, spinor fields, vector fields...) in the matter action? In this case, the Einstein equations together with the Bianchi identities would imply the conservation of the complete TDiff matter EMT, although not necessarily that of each TDiff field separately. Nevertheless, we would still obtain a consistency condition on the metric (note also that at no point along the reasoning had we the need to specify any coupling function $f_k$ nor $f_v$ for any of the fields: they have remained arbitrary all throughout and do not affect that conclusion). There is currently work being done in this direction in cosmological contexts, see for instance reference \cite{diego} for the study of two TDiff scalar fields and reference \cite{alfredo} for the study of a TDiff coupled vector field interacting with a Diff coupled spinor field.

Finally, a word on coupling functions. The TDiff scalar field model we consider in this work is also studied in references \cite{Alvarez:2009ga,maroto_tdiff_2023}. In those references it is argued that if one wishes the scalar field to preserve the Weak Equivalence Principle (WEP) then both functions should coincide, i.e. $f_k = f_v \equiv f$. Although this condition makes the scalar field model phenomenologically viable for matter in the visible sector, it may not be necessary in the dark sector. With this possible application in mind, then, we have in this work allowed both functions to differ from each other. But, in general, for any type of field in the geometric optics approximation, one recovers the standard Diff-invariant predictions when all the couplings are the same. It is therefore natural to expect that the WEP is satisfied when the new coupling functions coincide. Currently, there is work being done \cite{alfredo} in the direction of revising the hypotheses for the WEP when the model is not a scalar field but fields of other spins.

To conclude, the present work considers a TDiff scalar field as a particular case of study. We have maintained the situation as general as possible with the idea of gaining some intuition on the gravitational implications of TDiff invariant fields. Therefore, we hope that the results presented in this work could suggest interesting particular models to investigate in further detail as promising candidates for describing the dark sector.

\appendix

\section{Covariantized action}\label{section: Covariantized action}

Although a general action of the form
\begin{equation}\label{eq: general TDiff action}
    S_\text{TDiff}[g_{\mu\nu},\Psi] = \int d^4x\, \sum_i f_i(g) \, \Lagr_i\left( g_{\mu\nu}, \Psi,\partial_\mu\Psi \right)
\end{equation}
(with each $\Lagr_i$ a Diff scalar and $f_i(g)$ arbitrary coupling functions) breaks Diff invariance down to TDiff, it is possible to rewrite it in a generally covariant way by introducing a Diff scalar density $\bar{\mu}$ which transforms as $\sqrt{g}$ under general coordinate transformations. For instance, following \cite{Henneaux:1989zc,maroto_tdiff_2023}, let us define the density to be
\begin{equation}
    \bar{\mu} = \partial_\mu\left(\sqrt{g} \, T^\mu\right) \,,
\end{equation}
with $T^\mu$ a Diff vector field. The quantity $\bar{\mu}$ indeed transforms as $\sqrt{g}$, since
\begin{equation}
    \frac{\bar{\mu}}{\sqrt{g}} = \frac{1}{\sqrt{g}} \partial_\mu\left(\sqrt{g} \, T^\mu\right) = \nabla_\mu T^\mu \equiv Y
\end{equation}
(where we have introduced the Diff scalar field $Y \equiv \nabla_\mu T^\mu$) so that, rearranging,
\begin{equation}
    \bar{\mu} = Y \sqrt{g} \,.
\end{equation}
Consider now the following Diff invariant action
\begin{equation}
    S_\text{Diff}[g_{\mu\nu},\Psi,T^\mu] = \int d^4x\, \sum_i \sqrt{g} \, H_i\left( Y \right) \Lagr_i\left( g_{\mu\nu}, \Psi,\partial_\mu\Psi \right) \,,
\end{equation}
which, at face value, seems to be describing a similar theory to that of action \eqref{eq: general TDiff action}, but with an additional vector field $T^\mu$ field coming into play via the arbitrary functions $H_i(Y) = H_i(\nabla_\mu T^\mu)$. For convenience, let us rewrite the arbitrary functions $H_i(Y)$ as 
\begin{equation}
    H_i(Y) \equiv Y f_i(Y^{-2}) \,,
\end{equation}
with arbitrary $f_i$, so that the action is rewritten as
\begin{equation}
    S_\text{Diff}[g_{\mu\nu},\Psi,T^\mu] = \int d^4x\, \sum_i \sqrt{g} \,Y f_i(Y^{-2}) \Lagr_i\left( g_{\mu\nu}, \Psi,\partial_\mu\Psi \right) \,.
\end{equation}
Interestingly, it is in fact possible to make this Diff action completely equivalent to the TDiff action \eqref{eq: general TDiff action} above by going to a coordinate system where $\bar{\mu}=1$. Indeed, in such a coordinate system we have that
\begin{equation}
    \sqrt{g} \,Y f_i(Y^{-2}) \bigg|_{\bar{\mu}=1} = \bar{\mu} \, f_i\left(g/ \bar{\mu}^2\right) \bigg|_{\bar{\mu}=1} = f_i(g) \,,
\end{equation}
and so we immediately see that
\begin{equation}
    S_\text{Diff}[g_{\mu\nu},\Psi,T^\mu] \bigg|_{\bar{\mu}=1} = \int d^4x\, \sum_i f_i(g) \Lagr_i\left( g_{\mu\nu}, \Psi,\partial_\mu\Psi \right) = S_\text{TDiff}[g_{\mu\nu},\Psi]\,.
\end{equation}
As a concrete example, and one of particular interest for us, let us take the Diff invariant action
\begin{equation}\label{eq: covariantized model action}
    S_\text{Diff}[g_{\mu\nu},\psi,T^\mu] = \int d^4x\, \sqrt{g} \left[ H_k(Y) \frac{1}{2} g^{\mu\nu} \partial_\mu \psi \partial_\nu\psi - H_v(Y) V(\psi)  \right] \,.
\end{equation}
If we set $\bar{\mu}=1$ and recognize the functions $f_k(g)$ and $f_v(g)$, we immediately realize that it has become precisely the TDiff action \eqref{eq: model matter action}. In this manner, we may alternatively work within the TDiff approach or the covariantized approach.

\section{Calculation of the transverse constraint}\label{section: Calculation of the transverse constraint}
\noindent In this Appendix we derive the transverse constraint \eqref{eq: ligadura transversa}. Our starting point is equation \eqref{eq: starting point for transverse constraint}, which we rewrite here for convenience:
\begin{equation}
    \Vec{\nabla}\ln\abs{\frac{C_\psi \sqrt{g} (1-F)}{f_k \delta V^2}} = \frac{-1}{2(1-F)} \Vec{\nabla} \ln \abs{C_g \delta V^2}\,.
\end{equation}
In what follows, we shall abbreviate notation and assume that all logarithms have an absolute value sign included, i.e. $\ln x \equiv \ln\abs{x}$ for the following calculations. Expanding the above expression and rearranging some multiplicative factors, we obtain
\begin{equation}
    -2(1-F) \left[ \Vec{\nabla}\ln C_\psi + \Vec{\nabla}\ln(1-F) - \Vec{\nabla}\ln\left(\frac{f_k}{\sqrt{g}}\right) - \Vec{\nabla}\ln \delta V^2 \right] = \Vec{\nabla}\ln C_g + \Vec{\nabla} \ln \delta V^2 \,,
\end{equation}
and grouping like terms in the RHS it follows that
\begin{equation}
    -2(1-F) \left[ \Vec{\nabla}\ln C_\psi + \Vec{\nabla}\ln(1-F) - \Vec{\nabla}\ln\left(\frac{f_k}{\sqrt{g}}\right) \right] = \Vec{\nabla}\ln C_g + (2F-1) \Vec{\nabla} \ln \delta V^2 \,.
\end{equation}
Let us now focus on the LHS of the above equation, which we may write as
\begin{equation}
    \begin{split}
        -2(1-F)\Vec{\nabla}\ln C_\psi & - 2\Vec{\nabla}(1-F) +(1-F)(2F-1) \frac{\Vec{\nabla}g}{g} =\\
        &=-2(1-F)\Vec{\nabla}\ln C_\psi + (2F-1) \left[ \Vec{\nabla}\ln(2F-1) + (1-F) \frac{\Vec{\nabla}g}{g} \right]\,,
    \end{split}
\end{equation}
where we have made use of the fact that differentiating a constant yields zero, so that we could very well write
\begin{equation}
    -2\Vec{\nabla}(1-F) = 2\Vec{\nabla}F = \Vec{\nabla}(2F) = \Vec{\nabla}(2F-1) = (2F-1) \frac{\Vec{\nabla}(2F-1)}{(2F-1)} = (2F-1) \Vec{\nabla}\ln(2F-1) \,.
\end{equation}
Having rewritten the LHS, we may straightforwardly rearrange the result so that we obtain
\begin{equation}
    -2(1-F)\Vec{\nabla}\ln C_\psi = \Vec{\nabla}\ln C_g + (2F-1) \left[ \Vec{\nabla}\ln \delta V^2 - \Vec{\nabla}\ln(2F-1) - (1-F) \frac{\Vec{\nabla}g}{g} \right] \,.
\end{equation}
Next we focus on the square bracket on the RHS of the above expression, which using our longitudinal constraint will actually simplify greatly:
\begin{equation}
    \begin{split}
        \Vec{\nabla}\ln \delta V^2 - \Vec{\nabla}\ln(2F-1) - \frac{\Vec{\nabla}g}{g} + F \frac{\Vec{\nabla}g}{g} &= \Vec{\nabla}\ln \left( \frac{\delta V^2}{g(2F-1)} \right) + g \frac{f'_k}{f_k} \frac{\Vec{\nabla}g}{g} =\\
        &= \Vec{\nabla} \ln\left( \frac{1}{C_g f_k} \right) + \Vec{\nabla}\ln f_k = - \Vec{\nabla}\ln C_g \,.
    \end{split}
\end{equation}
Substituting this result back, we obtain
\begin{equation}
    -2(1-F) \Vec{\nabla}\ln C_\psi = 2(1-F) \Vec{\nabla}\ln C_g \,,
\end{equation}
which after simplification finally yields the transverse constraint
\begin{equation}
    \Vec{\nabla}\left( C_g C_\psi \right) = 0 \,.
\end{equation}

\acknowledgments
The authors would like to thank David Alonso López and Javier de Cruz Pérez for useful comments and discussions regarding TDiff theories. DJG also acknowledges financial help from the Ayudas de Máster IPARCOS-UCM/2022. This work has been supported by the MICIN (Spain) projects PID2019-107394GB-I00 and PID2022-138263NB-I00 (AEI/FEDER, UE).

\bibliographystyle{JHEP}
\bibliography{bibTDiff}

\end{document}